\documentclass[onecolumn,showpacs,preprintnumbers,amsmath,amssymb]{revtex4}

\newcommand{\al}{\alpha}

\newcommand{\ba}{\begin{align}}
\newcommand{\bpm}{\begin{pmatrix}}
\newcommand{\epm}{\end{pmatrix}}
\newcommand{\be}{\begin{equation}}
\newcommand{\ee}{\end{equation}}

\usepackage{graphicx}
\usepackage{dcolumn}
\usepackage{bm}
\usepackage{color}

\begin{document}


\title{Spacetimes foliated by non-expanding null surfaces  in the presence of a cosmological constant}

\author{Jerzy Lewandowski}
\author{Adam Szereszewski}%
\affiliation{Faculty of Physics, University of Warsaw, Pasteura 5, 02-093 Warsaw, Poland}


\date{\today}

\begin{abstract}

We prove that every   solution to Einstein's equations with possibly non-zero cosmological constant that is foliated by non-expanding null surfaces transversal
to a single non-expanding null surface belongs to family of the  near (extremal)  horizon geometries. Our results are local, hold in a neighborhood of the single
non-expanding null surface.

\end{abstract}

\maketitle

\section{\label{Intro}Introduction}   Spacetimes foliated by
non-expanding  null surfaces were introduced by Kundt  in the context
of  explicit ("exact") solutions to Einstein's equations describing
plane fronted gravitational waves \cite{Kundt}. In that framework the
primary structure is a congruence of null geodesic curves that is
non-diverging and surface forming \cite{PO}.  Another topic of General
Relativity in which non-expanding null surfaces appear is theory of
non-expanding horizons (NEHs) \cite{ABL,LewPawhigher}.  Within that
theory a geometric description of NEHs was developed, the constraints
following from the Einstein's equations satisfied by a surrounding
$n+2$ dimensional spacetime $M$ were studied.
Upon some energy assumptions that are satisfied by all the solutions of the vacuum Einstein's equations with a possibly non-zero cosmological constant considered in the current paper,  the  internal   geometric structure induced on  every NEH $H$ was identified as a degenerate metric tensor $g_{ab}$ of the signature $(0+...+)$ and a torsion free covariant derivative $\nabla_a$  such that 
\be\nabla_ag_{bc}=0.\ee
While every null vector field $\ell$ tangent to $H$ satisfies 
\be{\cal L}_\ell g_{ab}=0\ee
identically, the geometry of $H$ may admit a stronger symmetry  generator $\ell$ that
satisfies
\be[{\cal L}_\ell, \nabla_a]=0.\ee
In the latter case $H$ is called an isolated horizon.  A rotation $1$-form potential $\omega^{(\ell)}$ of a NEH $H$, depends on choice of a null vector $\ell$, and is defined by
\be\label{rot1form}
\nabla_a\ell = \omega^{(\ell)}_a\ell . 
\ee
It satisfies  the zeroth law of NEH mechanics   
\be\label{0law} 
\nabla_a\kappa^{(\ell)}={\cal L}_\ell \omega^{(\ell)}, \ \ \ \ \ \ \  \kappa^{(\ell)}:=\ell^a\omega^{(\ell)}_a. 
\ee
For every isolated horizon  
\be \kappa^{(\ell)}={\rm const}\ee
and the horizon is called extremal whenever
\be \kappa^{(\ell)}=0.\ee
A geometrical identity that connects the internal and external
geometry of an extremal isolated horizon $(\tilde H, \tilde\ell)$
turns into a constraint on data induced on an $n$-dimensional section
$S$ of $\tilde H$:
the metric tensor $g_{AB}$, its torsion free metric connection $\nabla^{(n)}_A$ and Ricci tensor $R^{(n)}_{AB}$, and the pullback of the rotation $1$-form $\omega^{(\tilde{\ell})}_A$ on the one hand, and the spacetime Ricci tensor  $R_{\alpha\beta}$ on the other hand, namely\cite{ABL,LewPawhigher}
\be\label{extr} 
-\nabla^{(n)}_{(A}\omega^{(\tilde{\ell})}_{B)} - \omega^{(\tilde{\ell})}_{A} \omega^{(\tilde{\ell})}_{B} +\frac{1}{2}R^{(n)}_{AB} = \frac{1}{2}R_{AB}.
\ee
That geometric point of view can also be applied to non-expanding null surfaces  
\be\mathbb{R}\ni u\mapsto H_u\subset M\ee 
foliating  a Kund's class spacetime $M$.  The existence of a foliation may be used to distinguish  a family of null vector fields in spacetime by 
\be\label{elldu} \ell_\mu = -\nabla_\mu u .\ee
For every lief $H_u$, the corresponding vector field $\ell^\mu$ is tangent, and satisfies 
\be \kappa^{(\ell)}=0.\ee
Since the function $u$ is defined up to transformations
\be u' = f(u)\ee
it follows that $\ell$ is defined up to a factor  $f'(u)$ constant  on every NEH  $H_u$. 
The most intriguing is a constraint implied by the existence of the foliation, and constraining  the data induced in every $n$-dimensional section of $H_u$, namely \cite{LewPawhigher}
 \be\label{fol} \nabla^{(n)}_{(A}\omega^{(\ell)}_{B)} - \omega^{(\ell)}_{A} \omega^{(\ell)}_{B} +\frac{1}{2}R^{(n)}_{AB} = \frac{1}{2}R_{AB}.\ee
The first, easy observation is, that given a solution $g_{AB},\ \omega^{(\ell)}_{A}$ of the foliation constraint (\ref{fol}) we can transform it
into a solution   $g_{AB},\ \omega^{(\tilde \ell)}_{A}$,
\be 
\omega^{(\tilde \ell)}_{A}:=-\omega^{(\ell)}_{A}
\ee
 of the extremal isolated horizon equation (\ref{extr}).
 Secondly, it turns out, that if  $H$ is just any NEH (no foliation assumption), transversal to an extremal isolated horizon $(\tilde H,\tilde \ell)$, and a null vector field $\ell$ tangent to  $H$ is normalized at the intersection $S$, such that
\be \tilde \ell^\mu\ell_\mu = {\rm const},\ee 
then the geometry of $H$ satisfies the foliation constraint (\ref{fol}) at $S$ (even if no foliation exists) \cite{PawLewJez, LewSzerWal1}.   Hence, the foliation constraint may be induced by a transversal isolated horizon. 

The second  observation led to the construction of  a family of  solutions to vacuum Einstein's equations in $4$-dimensional spacetime  foliated by bifurcated Killing horizons whose common part is an extremal Killing horizon  \cite{PawLewJez,LewSzerWal1}.  Every solution in that family is determined  by a metric tensor $g_{AB}$ and  a $1$-form $\omega_A$ defined on a $2$-dimensional manifold $S$ and satisfying the following equation 
\be\label{fol1} 
\nabla^{(2)}_{(A}\omega_{B)} - \omega_A\omega_B + \frac{1}{2}R^{(2)}_{AB}\ =\ 0 .
\ee  
 The corresponding metric tensor is defined on $M=S\times\mathbb{R}\times\mathbb{R}$ in coordinates $(x^\mu)=(x^A,v,u)$ as follows
\be\label{fol2}
 g_{\mu\nu}dx^\mu dx^\nu = g_{AB}dx^Adx^B - 2du\left(dv + 2v\omega +\frac{1}{2}v^2\left(\nabla_A\omega^A - 2\omega_A\omega^A  \right) du\right) .\ee   
The foliation consists of surfaces 
$$H_u\ :\ u=u_0, \ \ \ \ \ \ \ u_0\in\mathbb{R}, $$
the transversal extremal Killing horizon   is   the surface 
$${\tilde H}\ :\ v=0,$$ 
the extremal Killing vector is $\partial_u$ while the non-extremal Killing vectors are 
$v\partial_v -(u-u_0)\partial_u$.                                 
A spacetime (\ref{fol2}) is a generalization of the solutions obtained by the Bardeen-Horowitz  limit from extremal Kerr black hole solutions  and these days is known better as near horizon geometry \cite{BarHor,Real,KunLuc}. They were
generalized to higher dimensions, to $\Lambda\not=0$, and to various kinds of matter fields \cite{KunLuc,LewSzerWal1}.


\section{\label{NEHandIH}Kund's class spacetimes containing a transversal NEH}
The mechanism of inducing the foliation constraint (\ref{fol}) by existence of a transversal isolated horizon 
suggest  a natural characterization  of  Kund's class metric by  existence or not of a transversal extremal isolated horizon. In fact, a transversal NEH will automatically admit a null symmetry that will make it isolated horizon \cite{LewSzerWal1}. We follow that clue  and divide  the Kunds class spacetimes into two types:
\begin{enumerate}
\item[(i)]  containing  a transversal NEH (possibly, after the extension along the null generators of the lives of the foliation)
\item[(ii)] not containing any transversal NEH 
\end{enumerate}

In the current paper we   consider Kund's class spacetimes of the  type (i)  above. We find all the solutions to the vacuum Einstein equations with possibly non-zero cosmological constant in $4$-dimensional spacetime,
\be\label{E} R_{\mu\nu}=\Lambda g_{\mu\nu}.\ee
Our derivation applies also to arbitrary dimension $n\ge 4$ spacetime   if some generalization of the characteristic Cauchy problem is true there.   

\subsection{Example: NHG spacetime}
An example of a spacetimes we are looking for is the  metric tensor
defined on  
$$M=S\times\mathbb{R}\times\mathbb{R}$$ 
in coordinates system $(x^\mu)=(x^A,v,u)$ corresponding to the product as follows
\be\label{folprop}
 g_{\mu\nu}dx^\mu dx^\nu = g_{AB}(x)dx^Adx^B - 2du\left(dv + 2v\omega_A(x)dx^A +\frac{1}{2}v^2\left(\nabla_A(x)\omega^A(x) - 2\omega_A(x)\omega^A(x) - \Lambda \right) du\right),\ee   
where the $1$-form $\omega_A(x)$ and metric tensor $g_{AB}(x)$ defined on $S$ satisfy the foliation constraint 
\be\label{folLambda} 
\nabla^{(2)}_{(A}\omega_{B)} - \omega_A\omega_B + \frac{1}{2}R^{(2)}_{AB} - \frac{1}{2}\Lambda g_{AB}\ =\ 0 .
\ee  
The NEH foliation is given by 
\be H_u = S\times\mathbb{R}\times\{u\}, \ \ \ \ \ \ \ \ u\in \mathbb{R} .\ee 
The transversal NEH is 
\be \tilde{H} = S\times\{0\}\times\mathbb{R}. \ee
 The vector field 
\be \ell :=\partial_v \ee  
satisfies 
\be\ell_\mu = -\nabla_\mu u \ee
 as required. For every $u$, on the NEH $S\times\mathbb{R}\times \{u\}$, the rotation $1$-form  potential is 
\be \omega^{(\ell)} = \omega_A(x) dx^A . \ee
By inspection one can  check that (\ref{fol2}, \ref{folLambda}) satisfies the Einstein's equations 
(\ref{E}).

\subsection{Einstein spacetimes foliated by NEHs transversal to a single NEH} 
Consider a  general spacetime of the type (i) above, that is foliated by non-expanding null surfaces transversal to a single non-expanding null surface and assume it satisfies the Einstein's equations
\be  
R_{\mu\nu} \ =\ \Lambda g_{\mu\nu}.
\ee
Let 
\be \mathbb{R}\ni u\mapsto H_u \ee 
be the family of non-expanding null surfaces, and $\tilde H$ be the transversal non-expanding null surface.
Locally, in a suitable neighborhood of a point of $\tilde H$,  we choose the 
function $u$ to be one of spacetime coordinates.  A vector field $\ell^\mu$ defined by
\be \ell_\mu = -\nabla_\mu u \ee
is tangent to each of the leaves $H_u$, null, and satisfies
\be\label{lnablal} \ell^\mu\nabla_\mu\ell = 0.\ee
Define the second coordinate, a function $v$, such that 
\be \label{v} 
\ell^\mu\nabla_\mu v = 1, \ \ \ {\rm and}\ \ \ v_{|{\tilde H}}=0. 
\ee
Locally, for every point of $\tilde{H}$, we may consider a neighborhood in $M$, such that every $H_u$ and $\tilde H$  have the topology
 \be  H_u,  \tilde H = S\times \mathbb{R}\ee
 and the intersection is
 \be  H_u \cap  \tilde H = S, \ee
where $S$ is a $2$-dimensional manifold. 
The remaining two spacetime coordinates $x^A$, $A=1,2$  are introduced first on $S$,
next pass to $H_{0} \cap  \tilde H$, next extended  along $\tilde H$ such that they are constant
along the null generators, and finally along each $H_u$ by
\be \ell^\mu\nabla_\mu x^A=0 .  \ee 
In the resulting coordinates 
\be (x^\mu)=(x^A,v,u) \ee
the metric tensor $g$ takes the following form 
\be\label{gmunu} g_{\mu\nu}(x,v,u)dx^\mu dx^\nu = g_{AB}(x,v,u)dx^Adx^B - 2du\left(dv + W_A(x,v,u)dx^A + H(x,v,u)du\right), \ee 
where  the dependence on the coordinates $v$ and $u$ is restricted by the properties of non-expanding null surfaces. 
Indeed, to begin with
\be  g_{AB}(x,v,u)_{,v}=0, \ \ \ \ {\rm and}\ \ \ \ g_{AB}(x,0,u)_{,u}=0. \ee
Due to the Einstein's equations  (\ref{E}) the first equality follows from the non-expanding of each $H_u$,
and the second equality follows from the non-expanding of $\tilde H$. Consequently, 
\be   g_{AB}(x,v,u) =  g_{AB}(x)\ee
is a unique metric tensor on $S$. 

Secondly, the second assumption in (\ref{v}) about $v$ implies
\be\label{WHv0} W_A(x,0,u)=0 = H(x,0,u). \ee

Next, for every $H_u$, the rotation $1$-form potential is
\be  \omega^{(\ell)}_A=\frac{1}{2}W_ {A,v}.\ee
The $0$th law (\ref{0law}) and (\ref{lnablal}) imply
\be  W_{A,vv}(x,v,u) = 0. \ee
Hence,
\be \label{W_A} W_A(x,v,u)= 2v\omega_A(x,u) .\ee
One can also notice, that at the transversal non-expanding null surface $\tilde H$, 
as a tangent null vector field we can use 
$$\tilde \ell :=\partial_u.$$ 
Since by the construction
\be \tilde{\ell}^\mu\ell_\mu = -\partial_u u = -1\ee
the $1$-form $\omega_A$ is also related to the rotation $1$-form potential on $\tilde H$,
\be \omega^{(\tilde \ell)}(x,u) = \kappa^{(\tilde{\ell})}(x)du -\omega_A(x,u)dx^A.\ee
Of course the $0$th law (\ref{0law}) still applies, but at this point we do not know anything
else about the dependence of $u$ and $x$. 

In order to  show that $\omega_A(x,u)$ is independent of $u$, we will refer to  the  holographic 
characterization  of  a spacetime  that contains two transversal  NEHs \cite{Racz}.  Consider a $4$-dimensional spacetime $M$, equipped with a metric tensor $g_{\mu\nu}$ that satisfies
the vacuum Einstein's equations with  given cosmological constant $\Lambda$. Suppose $H\subset M$ and ${\tilde H}\subset M$ 
are two NEHs, and 
$$S=H\cap {\tilde H}$$
is a $2$-dimensional spacelike surface in $M$. Then, the metric tensor $g_{\mu\nu}$  in the future and in the past of $S$ is
locally, near $S$, and up to diffeomorphisms,  determined by the following holographic data induced on $S$: 
\begin{itemize}
\item the induced metric tensor $g_{AB}$,
\item the pullback $\omega_{AB}$ of the rotation $1$-form potential $\omega^{(\ell)}_a$ defined on $H$ by a tangent, null vector field 
$\ell$.
\end{itemize}
Moreover, two data sets,  $(g_1, \omega_1)$ and  $(g_2, \omega_2)$ related either by a diffeomorphism of $S$ or by 
$$\omega_2 = \omega_1+df, \ \ \ \ g_1=g_2 $$
define diffeomorphically equivalent metric tensors. 

We go back now to the metric tensor (\ref{gmunu}). Choose any value $u_0$
of the coordinate $u$ and consider  the holographic data $g_{AB}(x)$ and $\omega^{(\ell)}_A(x,u_0)$
defined on the intersection
$$S_{u_0}=H_{u_0}  \cap\ {\tilde H}=S\times\{0\}\times\{u_0\}.$$
Next we notice, that   among the  spacetimes (\ref{folprop}) there is one, that defines the same (up to diffeomorphisms) holographic data. To construct it, just set  in the definition of the metric tensor (\ref{folprop}) the following  
\be g_{AB}(x):=g_{AB}(x), \ \ \ \ \omega_A(x,u_0):= \omega^{(\ell)}_A(x,0,u_0), \ee
defined on $S$, considering  $u_0$ in $\omega_A$ as a parameter.  
Denote the resulting metric by $g_{(u_0)}$. The holographic data it defines on 
$$S_{u'_0}=H_{u'_0}\cap {\tilde H} = S\times\{0\}\times\{u'_0\}$$  
for arbitrarily fixed $u'_0$ is the same (up to the diffeomorphism) as the holographic data 
defined on $S_{u_0}$ by the metric $g$.   The conclusion is, that the metric tensor $g$ (\ref{gmunu}) 
restricted to the future and past of $S_{u_0}$ is isometric to the metric $g_{(u_0)}$ (\ref{folprop}) restricted to 
the future and the past of $S_{u'_0}$.  What may happen is, that the coordinates $(x^A,v,u)$ are not preserved
by that isometry. However they transform within the remaining freedom, 
\be 
u' = U(u), \qquad v' = \frac{v}{U'(u)}, \qquad  x'^A=x^A.
\ee
 Therefore without lack of generality, we change them in (\ref{gmunu}) such that actually the isometry
 preserves the coordinates. We can also fix $u'_0=0$. In those coordinates the components $W_A(x,v,u)$ 
 and $H(x,v,u)$ in (\ref{gmunu}) coincide with the corresponding components
 of the metric (\ref{folprop}) for $u,v>0$ and for $u,v<0$. In the case of $W_A$ it means that 
 \be 2v\omega_A(x,u)= 2v\omega_A(x), \ \ \ {\rm for}\ \ \ u,v>0, \ u,v<0 . \ee
 This is sufficient to conclude that 
 \be \omega_A(x,u)=\omega_A(x)\ee
 meaning that in (\ref{gmunu}), 
 \be\label{Wsol}W_A(x,v,u) = 2v\omega_A(x). \ee
 
 To determine the function $H(x,v,u)$ in (\ref{gmunu}), we just conclude from
 the comparison with (\ref{folprop}) in the domain $v,u<0$ or $v,u>0$, by the 
 second differentiability of $H$, that
 \be\label{Hv} H(x,0,u)_{,v}=0.\ee
 With the initial conditions (\ref{WHv0}) (the second equation) and (\ref{Hv}) at $v=0$, we can integrate 
 the Einstein's equation given by the component $R_{uv}$ of the Ricci tensor. We calculate
 $R_{uv}$ for the metric (\ref{gmunu}) with (\ref{Wsol}) (the first equation below) and use 
 the Einstein's equations (the second equation), 
 \be
      R_{uv} = - \nabla^{(n)}_A  \omega^A + 2 \omega^A\omega_A + H_{,vv} = -\Lambda.      \label{Ruv}
 \ee
The  result is
 \be
    H(x,v) = \frac{1}{2}\left(\nabla_A \omega^A - 2\omega_A\omega^A -\Lambda\right)v^2.    \label{H}
 \ee
 and as it could be anticipated, it coincides with (\ref{folprop}). 

\section{Summary}    
In summary,  we have proved the following 
\medskip

\noindent{\bf Theorem:} {\it Suppose $M$ is a $4$-dimensional spacetime equipped with 
a metric tensor $g$ that satisfies Einstein's equations
\be R_{\mu\nu}=\Lambda g_{\mu\nu},\ee
and admits a foliation by non-expanding null surfaces transversal to a single non-expanding
null surface $\tilde H$. Locally, in a neighborhood of every point of $\tilde H$, 
the metric $g$ can be written in the form (\ref{folprop}, \ref{folLambda}). 
}
\medskip

The resulting spacetime is the known near horizon geometry. It has $2$-dimensional family 
of Killing vectors, linear combinations of: 
$$K:=\partial_u, \ \ \ \ L:= u\partial_u-v\partial_v.$$ 
The non-expanding surfaces $u={\rm u_0}$ and $v=0$ are Killing Horizons 
of $ (u-u_0)\partial_u-v\partial_v$ while $v=0$ is at the same time the extremal Killing horizon
of $K$ \cite{MPS1,MPS2}. 

Our result generalizes easily to an arbitrary dimension $n$ provided it continues to be true
that a solution to the vacuum Einstein's equations with cosmological constant that admits
two intersecting generically non-expanding null surfaces $H$ and $H'$  is uniquely determined
(locally, in a neighborhood of the intersection) in the future and past of the intersection
by internal geometries $g_{ab}, \nabla_a$ and  $g'_{ab}, \nabla'_a$  induced on $H$, 
and, respectively $H'$. 
\medskip

\noindent{\bf Acknowledgments.}   
This work was partially supported by the Polish National
Science Centre\\ grant No. 2015/17/B/ST2/02871 .


\begin{thebibliography}{99}

\bibitem{Kundt}
W. Kundt,  The plane-fronted gravitational waves, Z. Physik {\bf 163}, 77--86 (1961).

\bibitem{PO}
J. Podolsk\'y and M. Ortaggio, Explicit Kundt type $II$ and $N$ solutions as gravitational
waves in various type $D$ and $O$ universes, Class. Quantum Grav. {\bf 20}, 1685--1701 (2003).

\bibitem{ABL}
A. Ashtekar, C. Beetle, and J. Lewandowski,
Geometry of Generic Isolated Horizon,
Class. Quantum Grav. {\bf 19}, 1195--1225 (2002), arXiv:gr-qc/0111067.

\bibitem{LewPawhigher}
 J. Lewandowski and T. Paw\l owski, Quasi-local rotating
black holes in higher dimension: Geometry, Classical
Quantum Gravity {\bf 22}, 1573 (2005).

\bibitem{PawLewJez} 
T. Paw\l owski, J. Lewandowski, and J. Jezierski,
Spacetimes foliated by Killing horizons,
Class. Quantum Grav. {\bf 21}, 1237--1252 (2004), arXiv:gr-qc/0306107.

\bibitem{LewSzerWal1}
J. Lewandowski, A. Szereszewski, and P. Waluk, Spacetimes foliated by non-expanding and Killing horizons: higher dimension, Phys. Rev. D {\bf 94}, 064018 (2016).



\bibitem{BarHor}
J.M. Bardeen and G.T. Horowitz,
Extreme Kerr throat geometry: A Vacuum analog of $AdS(2)\times S^2$,
Phys. Rev. D {\bf 60}, 104030 (1999), [arXiv:hep-th/9905099].

\bibitem{Real}
H. S. Reall, Higher dimensional black holes and supersymmetry, Phys. Rev. D {\bf 68}, 024024 (2003).

\bibitem{KunLuc} H.K. Kunduri and J. Lucietti,
\emph{Classification of Near-Horizon Geometries of Extremal Black Holes},
Living Rev. Rel. {\bf 16} (2013), 8,
http://www.livingreviews.org/lrr-2013-8, arXiv:abs/1306.2517.



\bibitem{Racz}  I. R\'acz, Stationary black holes as holographs II., Class. Quantum Grav. {\bf 31}, 035006 (2014).

\bibitem{MPS1}
Mars, M., Paetz, T.-T, and Senovilla, J.M.M., Multiple Killing Horizons, Class. Quantum
Grav. 35 (2018) 155015; arXiv:1803:03054

\bibitem{MPS2}
M. Mars, T.-T Paetz, and J.M.M. Senovilla, Multiple Killing Horizons and Near Horizon Geometries, arXiv:1807.02679.

\end{thebibliography}
\end{document}